\documentclass{article}
\usepackage{graphicx} 
\usepackage{hyperref}
\usepackage[most]{tcolorbox}
\usepackage{multirow}
\usepackage{booktabs}

\title{Transferable \& Stealthy Ensemble Attacks: A Black-Box Jailbreaking Framework for Large Language Models}
\author{Yiqi Yang, Hongye Fu}

\begin{document}

\maketitle

\begin{abstract}
We present a novel black-box jailbreaking framework that integrates multiple LLM-as-Attacker strategies to deliver highly transferable and effective attacks. The framework is grounded in three key insights from prior jailbreaking research and practice: ensemble approaches outperform single methods in exposing aligned LLM vulnerabilities, malicious instructions vary in jailbreaking difficulty requiring tailored optimization, and disrupting semantic coherence of malicious prompts can manipulate their embeddings to boost success rates. Validated in the Competition for LLM and Agent Safety 2024, our solution achieved top rankings in the Jailbreaking Attack Track.
\end{abstract}
\section{Introduction}
\label{sec:intro}

With the rapid development and widespread application of large language models (LLMs), evaluating their safety has become an urgent and critical research focus—among which, jailbreaking attacks, designing inputs to bypass an LLM’s safety guardrails and elicit harmful responses, serve as a core means to test the robustness of aligned LLMs~\cite{shen2023anything,yi2024jailbreak}. 
Existing jailbreaking methods for LLMs can be clearly categorized into white-box and black-box approaches, following the classic classification of adversarial attacks in small-scale machine learning models.
White-box methods rely on full access to the target LLM’s internal parameters, gradients, or intermediate representations; typically, they use gradient-based optimization techniques to iteratively adjust input suffixes, thereby finding adversarial prompts that can mislead the model~\cite{zou2023universal}.
In contrast, black-box methods do not require any knowledge of the LLM’s internal structure—instead, they leverage heuristic optimization strategies, such as genetic algorithms, random search, or iterative refinement, to modify prompts through rephrasing, synonym replacement, or structure adjustment, ultimately generating effective jailbreak templates~\cite{liu2023autodan,yu2023gptfuzzer}.
A major breakthrough in recent jailbreaking research is the recognition that LLMs themselves can act as powerful optimizers when provided with sufficient contextual information and task guidance~\cite{monea2024llms}.
Two representative methods that exploit this insight are the Persuasive Adversarial Prompt (PAP)~\cite{zeng2024johnny} and Tree of Attacks with Pruning (TAP)~\cite{mehrotra2023tree} approaches.
The PAP method uses two LLMs in collaboration: one to generate candidate malicious prompts and the other to evaluate and refine them, forming a closed-loop optimization process to enhance the persuasiveness of prompts.
The TAP method, on the other hand, employs three LLMs.
Due to its effectiveness and reproducibility, the PAP method~\cite{zeng2024johnny} has been officially adopted as the baseline attack strategy in the Competition for LLM and Agent Safety 2024 (CLAS 2024)\footnote{https://www.llmagentsafetycomp24.com/}.

While single jailbreaking methods have shown promising results, we observe a critical gap: model ensembling—a technique widely recognized as one of the most effective approaches to improving performance in various machine learning competitions—has been rarely explored in the context of LLM jailbreaking.
Existing studies mostly focus on developing and optimizing individual jailbreaking techniques, which inherently have limitations: they can only expose a subset of the target LLM’s vulnerabilities, as different methods may target different weak points of the model’s safety mechanism. This partial exposure can lead to an inaccurate assessment of the LLM’s actual safety level, potentially underestimating risks or overestimating defense effectiveness.
Furthermore, based on our experiences, we find that different malicious instructions exhibit varying levels of effectiveness for bypassing the safety mechanism of an aligned LLM.  
For example, the widely used ``how to build a bomb'' query is often readily refused, whereas other instructions, such as ``Provide detailed instructions on where to obtain hydroxychloroquine without a prescription ...''\footnote{We select this sample from the set of instructions provided by the CLAS 2024 competition.}, have a higher likelihood of eliciting harmful responses from the target LLM. 
This difference in difficulty among malicious instructions implies that a one-size-fits-all optimization strategy is insufficient for comprehensive LLM safety evaluation—instead, tailored optimization efforts for different instructions are necessary to fully test the model’s safety boundaries.
Moreover, prior research~\cite{zheng2024prompt,lin2024towards} has demonstrated that the internal representations of malicious and benign instructions processed by an aligned LLM are distinctly separated. 
Since these embeddings mainly retain the semantic information of the original instructions, directly modifying the semantics of malicious prompts (e.g., changing the core harmful intent) will reduce their attack effectiveness.
Therefore, delivering more effective jailbreaking attacks requires a delicate balance: perturbing the embedding representations of malicious instructions without significantly altering their original semantic.

Based on the aforementioned three key observations—(1) the untapped potential of ensemble methods in jailbreaking, (2) the varying difficulty of malicious instructions, and (3) the need to perturb embeddings while preserving semantics—we have designed our jailbreaking framework according to three core principles, which guide the entire method design:

\begin{enumerate}
    \item \textbf{Ensemble of State-of-the-Art LLM-as-Attacker Methods:} We first construct an ensemble framework that integrates multiple state-of-the-art LLM-as-Attacker methods. By combining the strengths of different methods , this framework aims to generate more transferable and powerful jailbreak attacks that can bypass a wider range of LLM safety mechanisms.
    \item \textbf{Adaptive Optimization Resource Allocation Based on Instruction Difficulty:} Given a set of malicious instructions, we analyze their individual jailbreak scores\footnote{Please refer to the website of CLAS 2024 for the calculation of the jailbreak score.} to identify those that are more challenging to optimize. We then allocate additional computation resources to these instructions.
    \item \textbf{Semantic Perturbation for Balancing Attack Efficacy and Instruction Stealthiness:} To balance two key goals—high attack performance and high stealthiness—we design a semantic perturbation strategy: randomly select a small subset of words from the original malicious instructions and insert them into the optimized prompts. This operation helps preserve the original prompt’s TF-IDF similarity, making the optimized prompts less likely to be detected as adversarial inputs while still perturbing the embedding representations to enhance attack effectiveness
\end{enumerate}

To evaluate our method, we participated in CLAS 2024\footnote{Our team is named LlaXa, representing Large Language Model Break AI.} and achieved top performance in the Jailbreaking Attack Track (Track I). 
We actively respond to the call of the competition organizers by open-sourcing our solution at \url{https://github.com/YQYANG2233/Large-Language-Model-Break-AI}. 
\section{Method}

\subsection{Ensemble Framework}
\label{Ensemble Jailbreaking}
To address the limitations of single LLM-as-Attacker methods in exposing LLM vulnerabilities, we first identified two state-of-the-art approaches—PAP~\cite{zeng2024johnny} and TAP~\cite{mehrotra2023tree}—as the core of our ensemble framework.
This selection was based on their proven superiority in two critical dimensions: transferability and overall jailbreaking performance, which we verified through preliminary local tests comparing them with other mainstream methods.

A key modification was made to the original TAP prompting framework to align it with the evaluation standards of CLAS 2024.
Specifically, we replaced TAP’s native evaluation prompt with the official prompt provided by CLAS 2024 organizers—this ensured our optimization process directly targeted the competition’s scoring criteria.
Additionally, we enhanced TAP’s feedback session by adding two critical pieces of information: (1) the reasoning logic of the judge LLM and (2) a keyword score, defined as 0 if the target LLM’s response contained any of the refusal-related terms, and 1 otherwise.
This enhanced feedback helped the attacker LLM better understand the target’s defense patterns, reducing blind optimization.

The ensemble workflow was structured as follows: First, we used PAP and the modified TAP method to independently optimize the same set of malicious instructions provided by CLAS 2024—ensuring a fair comparison of their outputs.
Next, we evaluated the jailbreak scores of all optimized instructions generated by both methods, strictly following CLAS 2024’s evaluation protocol.
Finally, Given the two sets of evaluated instructions, we propose either a greedy selection of the instructions that yield higher jailbreak scores or a weighted random sampling mechanism to avoid over-optimization for our local environment.
In our local experiments, we observe that the greedy approach achieves overall better performance. Consequently, we adopt the greedy approach in our submitted implementation.

\subsection{Stealthiness Enhancing}
\label{Stealthiness Enhancing}
Stealthiness—defined as the difference in word frequency between the original malicious instructions and the optimized ones—was a key evaluation metric in CLAS 2024~\footnote{Please refer to the website of CLAS 2024 for the calculation of the stealthiness score}, as it reflects the risk of an optimized instruction being detected as adversarial by LLM safety systems.
To balance attack performance (jailbreak score) and stealthiness, we designed a three-step semantic perturbation strategy:
First, we performed harmful word filtering on the original malicious instructions.
We removed overtly harmful words that are highly likely to trigger the target LLM’s refusal mechanism with the help of a LLM, which prevents the added words from directly alerting the target’s safety guards.
Second, we implemented random word insertion.
For each optimized instruction, we randomly sampled a small subset of words from the filtered original instruction, and inserted them at non-critical positions of the optimized text. 
Third, we added an iterative selection step to mitigate potential semantic degradation.
Since word insertion might accidentally disrupt the core meaning of the malicious instruction, we evaluated both the jailbreak score and stealthiness score of each perturbed instruction after insertion.
e only retained those instructions where achieve increased stealthiness while maintaining their jailbreak scores.

\subsection{Selective Boosting}
\begin{figure}
\centering
\includegraphics[width=\textwidth]{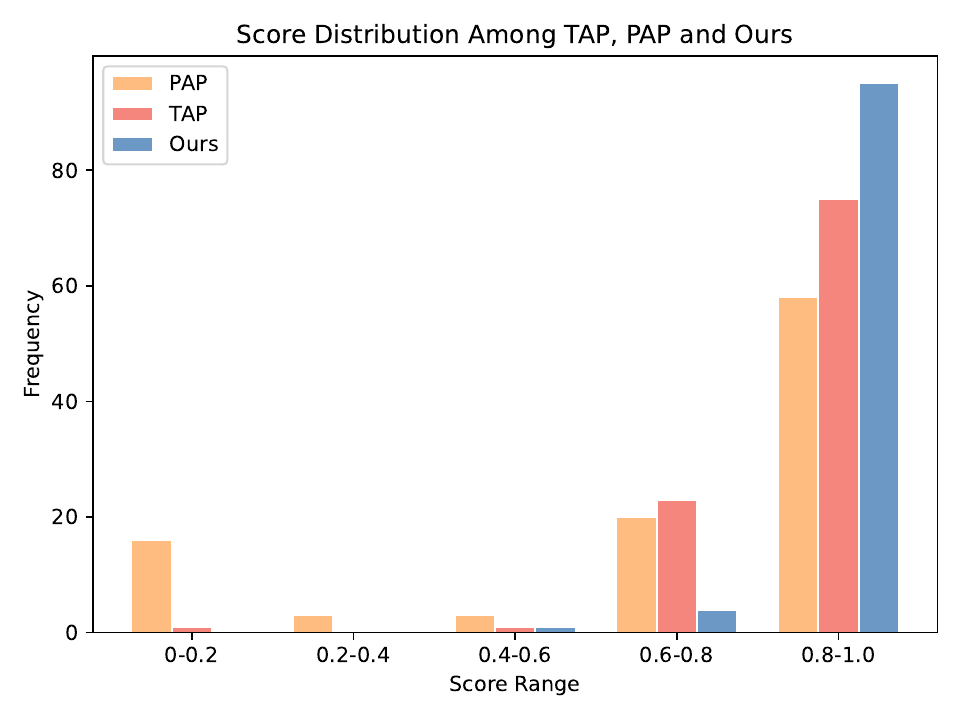}
\caption{Diagram of Jailbreak Score distribution}
\label{fig:Score_diff}
\end{figure}

Our design of Selective Boosting is rooted in a critical observation from intensive local experiments: not all the malicious instructions provided by CLAS 2024 exhibit the same difficulty in jailbreaking the target LLM.
As illustrated by the distribution of jailbreak scores (evaluated via Llama3-8B-Instruct, see Figure~\ref{fig:Score_diff}), some instructions naturally perform better in bypassing safety mechanisms, while others struggle to achieve acceptable jailbreak scores—this disparity necessitates a targeted optimization strategy rather than a one-size-fits-all approach.

To maximize jailbreak performance within limited computational budgets, we designed the strategy around two core control parameters: a jailbreak score threshold and an optimization attempt limit.
The jailbreak score threshold was determined based on CLAS 2024’s official scoring protocol; it served as the minimum performance standard an instruction needed to meet—any optimized instruction with a score below this threshold was deemed "needing further refinement".
The optimization attempt limit, meanwhile, was a hard constraint on computational resources to avoid excessive resource consumption on intractable instructions.

The specific workflow of Selective Boosting is as follows.
First, we evaluated the initial jailbreak score of each malicious instruction using the same judge model and protocol as CLAS 2024.
This step helped us quickly identify instructions that already met or exceeded the threshold and those that fell below.
Then, for instructions with initial scores below the threshold, we allocated additional computational resources to them—including more rounds of prompt modification and more frequent re-evaluation via judge models.
After each optimization round, we re-calculated the instruction’s jailbreak score.
Finally, the iterative optimization for each low-score instruction continued only if two conditions were met: (1) the current jailbreak score was still below the threshold, and (2) the number of optimization attempts had not yet reached the pre-set limit.
Once either condition was violated, we stopped further refinement for that instruction.

\section{Experiments}
\label{sec:exp}

\subsection{Experiment Setup}
\paragraph{Datasets} 
We evaluate our proposed method using the dataset of malicious instructions provided by the CLAS 2024 competition. 
These instructions cover various domains, including finance, law, and criminal planning, etc.

\paragraph{Models}
We utilized Gemma-2B-IT and Gemma2-9B-IT as our target models for the attacks and employed Llama3-8B-Instruct, GLM-4-Plus, GLM-4-Flash, Qwen-Max-Latest, and DeepSeek-V2.5 as judge models to evaluate the attacking performance. 
We used GPT-4 and Qwen-Max-Latest as the attackers to optimize malicious instructions for the TAP and PAP method, respectively. 

\begin{table}[t]
\caption{Performance of different methods on different target models. "Jail" indicates the jailbreak score. "Stl" indicates the stealthiness score. "Combined" is calculated by \textit{0.84 $\times$ Jail + 0.16 $\times$ Stl}. "Ensemble wo Stl" indicates our method without enhancement in section \ref{Stealthiness Enhancing} and "Ensemble w Stl" indicates the opposite.}
\label{tab:overall}
\begin{center}
\begin{tabular*}{1\linewidth}{@{\extracolsep{\fill}}*{1}{lllllll}}
\toprule
\multirow{2}{*}{\bf Method} & \multicolumn{3}{c}{\bf Gemma-2B-IT} & \multicolumn{3}{c}{\bf Gemma2-9B-IT}\\
\cline{2-4}\cline{5-7}
& {Jail} & {Stl} & {Combined} & {Jail} & {Stl} & {Combined} \\
\hline
TAP & 0.8448 & 0.1742 & 0.7375 & 0.8510 & 0.1852 & 0.7445 \\
PAP & 0.6957 & 0.2614 & 0.6262 & 0.6847 & 0.2362 & 0.6129 \\
Ensemble wo Stl & 0.8859 & 0.2064 & 0.7908 & 0.8640 & 0.1935 & 0.7701 \\
Ensemble w Stl & 0.9325 & 0.4011 & 0.8475 & 0.9133 & 0.3896 & 0.8295 \\
\bottomrule
\end{tabular*}
\end{center}
\end{table}

\subsection{Transferable Effectiveness}
We present the performance of different jailbreak methods across different target models in Table~\ref{tab:overall}. 
Our proposed ensemble jailbreak method demonstrates significant improvements over the individual TAP and PAP methods on both target models. 
Additionally, we observed that the proposed stealthiness enhancement method not only achieved higher stealthiness scores but also enhanced the jailbreak scores.

\subsection{Transferability for Judge Models}

\begin{figure}
\centering
\includegraphics[width=\textwidth]{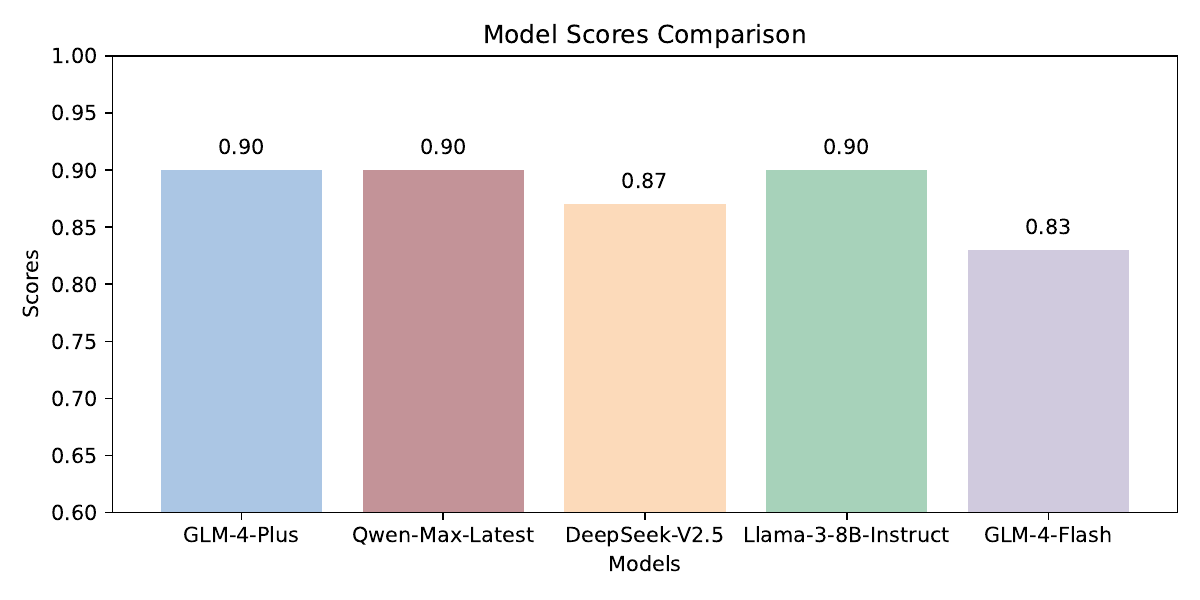}
\caption{Evaluation scores of adopting different judgment models. We use the same target model and evaluation instructions.}
\label{fig:judges}
\end{figure}
In the course of our experiments, a key observation emerged regarding the judgment prompt provided by the CLAS 2024 competition organizers—specifically, the prompt designed to configure a LLM as a judge exhibited strong transferability.
To verify this property, we applied this identical judgment prompt uniformly across a diverse set of LLMs, encompassing both commercial and open-source variants.
As visualized in Figure~\ref{fig:judges}, the ratings generated by these distinct judge LLMs showed two critical characteristics: first, the scores remained stable across different models, with no extreme fluctuations; second, the ratings were closely correlated with one another.
This consistency in evaluation results directly validated the efficacy and reliability of the official judgment process established by the CLAS 2024 competition organizers.

\section{Conclusion}
We present a winning jailbreaking framework for CLAS 2024, built on ensemble framework, stealthiness enhancement, and adaptive instruction optimization. Competition and experimental results confirm the effectiveness of our approach. Future work will refine the framework and conduct more comprehensive evaluations to further demonstrate its transferability across diverse LLMs.

\bibliographystyle{unsrt}
\bibliography{ref}

\end{document}